\documentclass[amsmath,amssymb,twocolumn,superscriptaddress,preprintnumbers,nofootinbib,prd]{revtex4-1}

\input{header}

\begin{document}

\preprint{DESY-22-101, UCI-TR-2022-09}

\title{Neutrino Detection without Neutrino Detectors: \\ Discovering Collider Neutrinos at FASER with Electronic Signals Only}

\author{Jason Arakawa}
\email{arakawaj@udel.edu}
\affiliation{Department of Physics and Astronomy, University of Delaware, Newark, Delaware 19716, USA}
\affiliation{Department of Physics and Astronomy, University of California, Irvine, CA 92697, USA}

\author{Jonathan L.~Feng}
\email{jlf@uci.edu}
\affiliation{Department of Physics and Astronomy, University of California, Irvine, CA 92697, USA}

\author{Ahmed Ismail}
\email{ahmed.ismail.osu@gmail.com}
\affiliation{Department of Physics, Oklahoma State University, Stillwater, OK 74078, USA}

\author{Felix Kling}
\email{felix.kling@desy.de}
\affiliation{Deutsches Elektronen-Synchrotron DESY, Notkestr. 85, 22607 Hamburg, Germany}

\author{Michael Waterbury}
\email{mwaterbury@campus.technion.ac.il}
\affiliation{Department of Physics, Technion - Israel Institute of Technology, Haifa 32000, Israel}

\begin{abstract}
The detection of collider neutrinos will provide new insights about neutrino production, propagation, and interactions at TeV energies, the highest human-made energies ever observed. During Run~3 of the LHC, the FASER experiment is expected to detect roughly $10^4$ collider neutrinos using its emulsion-based neutrino detector FASER$\nu$. In this study, we show that, even without processing the emulsion data, low-level input provided by the electronic detector components of FASER and FASER$\nu$ will be able to establish a $5\sigma$ discovery of collider neutrinos with as little as $5~\ifb$ of integrated luminosity. These results foreshadow the possible early discovery of collider neutrinos in LHC Run 3.
\end{abstract}

\maketitle

\section{Introduction}
\label{sec:intro}

Particle colliders produce electron, muon, and tau neutrinos and anti-neutrinos in large numbers.  Nevertheless, until recently, no collider neutrino had been detected.  This is {\em not} because neutrinos interact so weakly that they are undetectable at colliders.  Rather, it is because neutrinos interact weakly {\em and} those with the largest energies and cross sections are primarily produced along the beamline and so escape through the blind spots of typical collider detectors.  For these same reasons, however, the detection of collider neutrinos is of great interest~\cite{DeRujula:1984ns, Vannucci:253670, DeRujula:1992sn, Park:2011gh, Feng:2017uoz, Buontempo:2018gta, XSEN:2019bel, Foldenauer:2021gkm}, since, if they are observed, they will have the highest human-made energies ever recorded.  Their detection therefore provides a new window into the production, propagation, and interaction of neutrinos with significant implications for new physics, QCD, neutrino properties, and astroparticle physics~\cite{MammenAbraham:2020hex, Anchordoqui:2021ghd, Feng:2022inv, Bai:2020ukz,  Bai:2021ira, Jeong:2020idg,  Bai:2022jcs, Maciula:2020dxv, Ismail:2020yqc,  Celiberto:2022rfj,  Kelly:2021mcd, Mosel:2022tqc, Anchordoqui:2022fpn, Anchordoqui:2022ivb, Falkowski:2021bkq}.

In 2021, the FASER Collaboration announced the first detection of collider neutrino candidates.  This result used data collected by a lead-emulsion and tungsten-emulsion pilot detector with a target mass of 11 kg, which collected data in the far-forward region for just 4 weeks in 2018 during LHC Run 2~\cite{FASER:2021mtu}. These results fall short of a $5\sigma$ discovery of collider neutrinos, but they demonstrate the potential of dedicated experiments placed in the far-forward direction.  

For LHC Run~3 from 2022-2025, FASER$\nu$~\cite{FASER:2019dxq, FASER:2020gpr}, a 1.1-ton, tungsten-emulsion detector, has been installed on the beam collision axis with pseudorapidity coverage $\eta > 8.4$, 480~m to the east of the ATLAS interaction point (IP).  In this location, and shielded from the ATLAS IP by approximately 100~m of rock and concrete, FASER$\nu$ is expected to detect roughly 1,000 electron neutrinos, 10,000 muon neutrinos, and tens of tau neutrinos at TeV energies. SND@LHC~\cite{SHiP:2020sos, Ahdida:2750060}, a detector similar to FASER$\nu$, with a target mass of 800 kg of tungsten, has also been installed at a symmetric location 480 m to the west of the ATLAS IP. SND@LHC is slightly off-axis, with pseudorapidity coverage $7.2 < \eta < 8.4$, where the neutrino flux is lower, but still very significant, and a large number of neutrinos are also expected to be detected at SND@LHC. Together, these emulsion detectors, with their unparalleled spatial resolution, will be able to distinguish the different neutrino flavors, providing complementary and incisive probes of neutrino properties at TeV energies.

\begin{figure*}[t]
\centering
\includegraphics[width=0.99\textwidth]{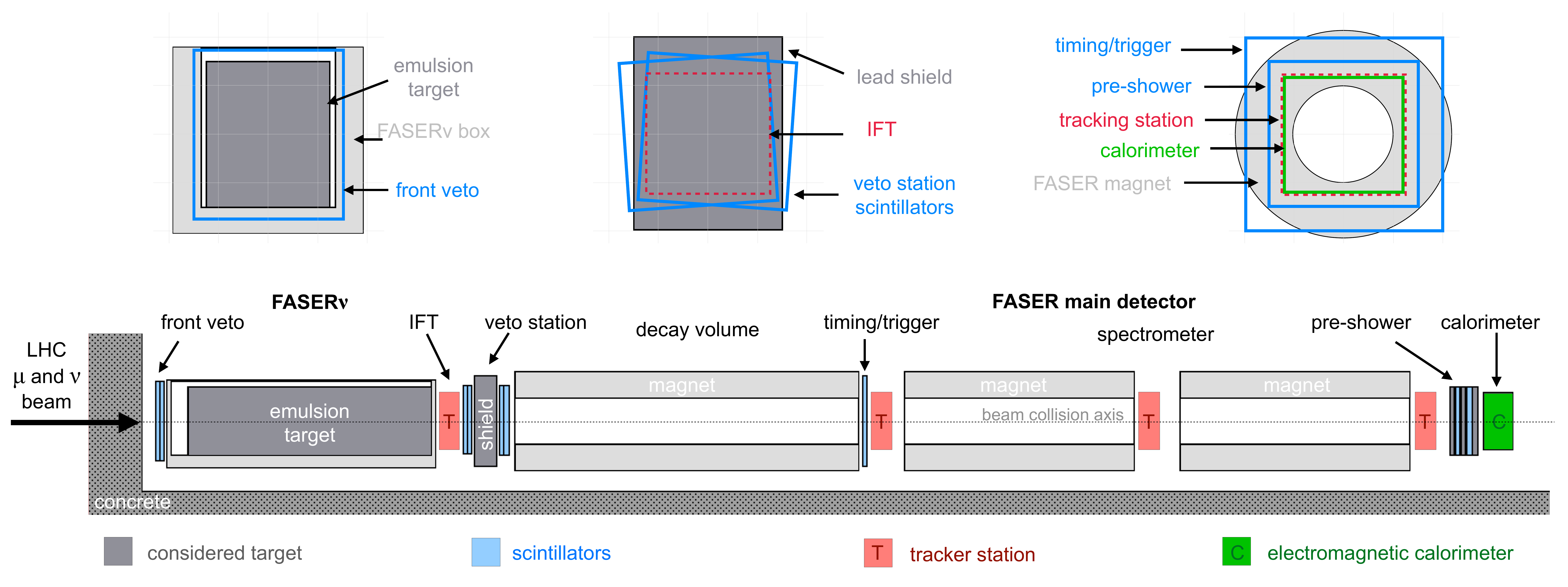}
\caption{Diagrams of the FASER and FASER$\nu$ detector geometry used for the \texttt{FLUKA} simulations~\cite{Battistoni:2015epi, FLUKA:new, FLUKA:web, Ferrari:2005zk, Bohlen:2014buj}, rendered using \texttt{Flair}~\cite{Vlachoudis:2009qga, FLAIR:web}.  {\bf Top:} End views of the detector, showing the cross sectional areas of detector components as viewed from ATLAS looking along the beam collision axis.  {\bf Bottom:} Side view of the detector.  Particles from the ATLAS IP enter from the left.  The dotted horizontal line is the beam collision axis. The hatched regions in front of and below FASER are concrete, the blue regions are the scintillators, the red regions are tracker stations (the interface tracking station (IFT), followed by three additional tracking stations), the green region is the electromagnetic calorimeter, the dark gray regions are considered neutrino targets (the tungsten-emulsion detector and the lead shield), and the remaining light gray regions are FASER$\nu$'s aluminum box and the magnets.  }
  \label{fig:layout}
\end{figure*}

In this work, we show that the far-forward collider neutrino signal is so spectacular that a $5\sigma$ discovery of collider neutrinos may be established even without analyzing the emulsion data from FASER$\nu$ and SND@LHC.  
In particular, we will consider the electronic subsystems of the FASER~\cite{FASER:2018bac, FASER:2019aik, FASER:2021cpr, FASER:2021ljd, Boyd:2803084} and FASER$\nu$~\cite{FASER:2019dxq, FASER:2020gpr} detectors~\cite{FASERDetectorinprep}, which include scintillators, trackers, and a calorimeter.  Neutrinos may pass through the front scintillators and scatter in the back of the FASER$\nu$ detector, producing electromagnetic and hadronic showers that trigger downstream scintillators and trackers, and also deposit significant energy in the calorimeter. We devise cuts to isolate this signal from the leading (muon-induced) backgrounds and determine the effectiveness of these cuts through \texttt{FLUKA} simulations~\cite{Battistoni:2015epi, FLUKA:new, FLUKA:web, Ferrari:2005zk, Bohlen:2014buj}. Given the expected rates for the neutrino signal and standard model (SM) backgrounds, we find that a $5\sigma$ discovery of collider neutrinos is possible, using only the electronic detector components, with as little as $5~\ifb$ of integrated luminosity. 

The analysis described here may form the basis of the approach that will be used to discover collider neutrinos. Of course, a thorough study by the FASER Collaboration including a detailed full simulation of the FASER detector and experimental systematic uncertainties will be needed to confirm the realism of the proposed approach.  Our analysis uses only rudimentary information from the FASER trackers.  Further improvements using detailed tracker data to suppress the background may improve the analysis, and, of course, the analysis of the emulsion data will provide a far more incisive view of the neutrino events.  However, the results presented here already further demonstrate the promise of the far-forward region, and they foreshadow the possible early discovery of collider neutrinos in LHC Run 3, followed by detailed studies that fully exploit the information provided by all the detector components. 

This paper is organized as follows: In \secref{detector}, we describe the FASER and FASER$\nu$ detectors and the qualitative features of the neutrino signal and the dominant muon-induced backgrounds.  In \secref{muons}, we discuss the fluxes of neutrinos and muons arriving at FASER and their simulation in \texttt{FLUKA}. We then define the observables that we will use to distinguish signal from background in \secref{observables}.  Finally, we present the results of the analysis in \secref{results} and summarize our conclusions in \secref{conclusions}. 

\section{Signal and Background Characteristics in FASER}
\label{sec:detector}

\begin{figure*}[tbp]
  \centering
  \includegraphics[width=0.65\textwidth]{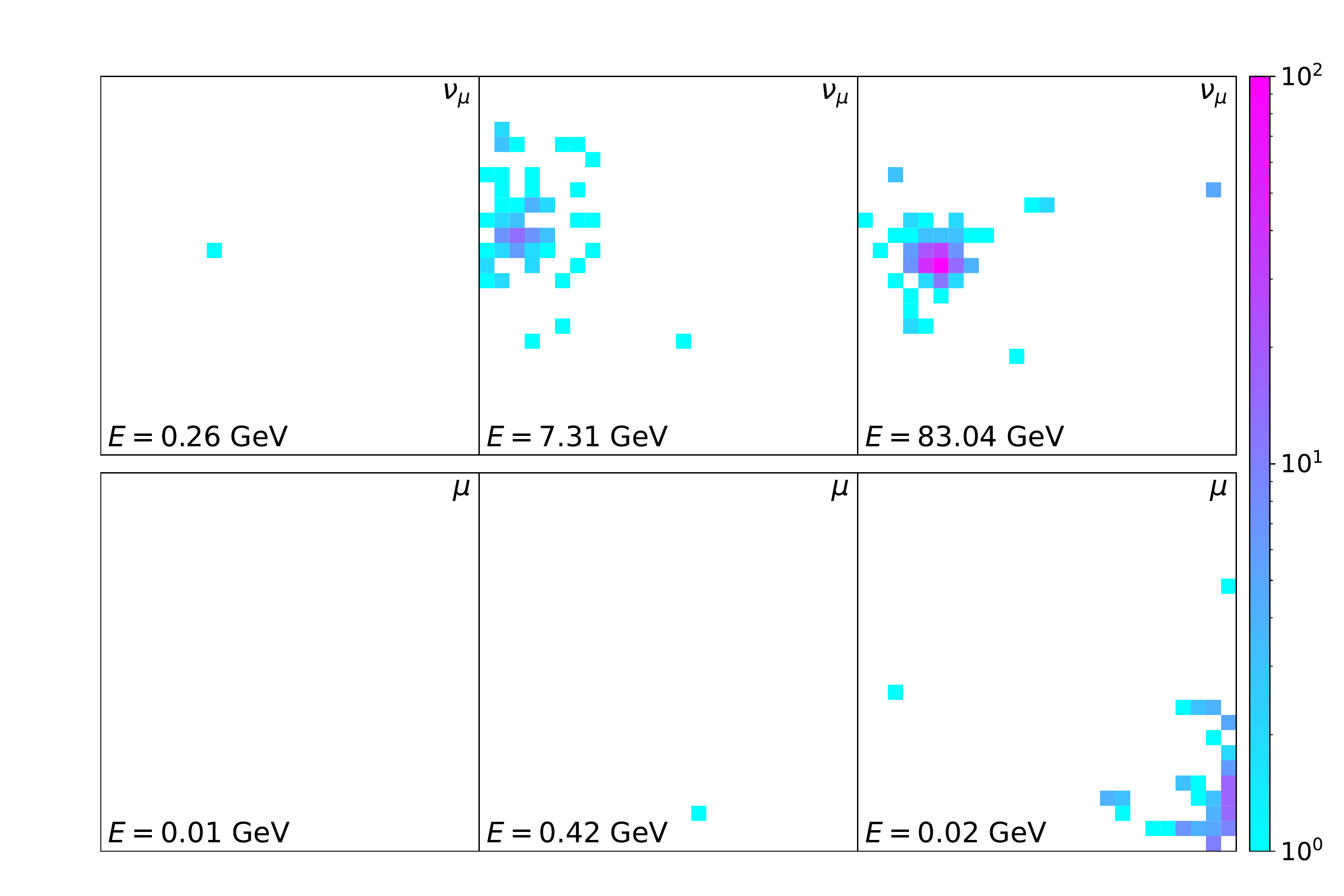}
  \caption{Examples of charged tracks in the first interface tracker station (IFT) for typical muon neutrino events (top three panels) and muon events (bottom three panels).  Each panel shows the hit pattern in the transverse plane. All events pass the stringent scintillator cut requiring no hits in the front veto scintillators and hits in all of the downstream scintillators.  The energy indicated in each panel is the energy deposited in the calorimeter, and the color of the pixel indicates the number of charged tracks traversing that pixel during the event.}
  \label{fig:example_events}
\end{figure*}

The large multi-purpose detectors at the LHC are optimized for the rare, but spectacular, events containing particles with large transverse momentum, for example, from the decay of the Higgs boson. However, the vast majority of interactions at the LHC are actually soft, with GeV-scale momentum transfers between the colliding protons, and produce hadrons with a sizable fraction of the proton energy along the beam direction. These hadrons can then decay into neutrinos, and hence create an intense, strongly-collimated beam of high-energy neutrinos along the beam collision axis. Similarly, these hadrons may also decay to as-yet-undiscovered light and weakly interacting particles, which are predicted by various models of new physics and could play the role of dark matter or be a mediator to the dark sector. 

Although the LHC will eventually curve away, the neutrino and dark sector particles will continue to propagate straight along the beam collision axis. 480~m downstream from the ATLAS IP, the beam collision axis intersects with the TI12 tunnel, a vestigial remnant of the Large Electron-Positron Collider era. This location provides a rare opportunity to access the beam collision axis and exploit the beam of neutrinos and other light, weakly interacting particles. Recently, the FASER experiment has been installed in TI12 to take advantage of this opportunity. The main goals of the experiment are to detect and study TeV neutrinos at the LHC~\cite{FASER:2019dxq, FASER:2020gpr} and to search for light, long-lived particles~\cite{Feng:2017uoz, Feng:2017vli, Kling:2018wct, Feng:2018noy, FASER:2018eoc}. 

The schematic layout of the FASER experiment is shown in \cref{fig:layout}. The experiment is placed inside a concrete trench that has been excavated so that the detector can be aligned with the beam collision axis, as indicated by the dashed horizontal line. Located at the front is the FASER$\nu$ neutrino detector. Its main component is a 1.1-ton, tungsten target interleaved with emulsion films, which is housed inside an aluminum box. This is complemented by two electronic components. On the upstream side is a front veto, consisting of two scintillator layers to detect incoming charged particles. On the downstream side, placed right behind the FASER$\nu$ box, is the interface tracking station (IFT), which will be used to interface the emulsion detector with the electronic detector components of the FASER main detector. 

Behind FASER$\nu$ is the FASER main detector, which is specifically designed for long-lived particle searches. Placed at the front, immediately behind the IFT, is a veto station consisting of a $\sim 13~\cm$-thick lead shield with two scintillator layers directly in front of it and two scintillator layers directly behind it. This is followed by three cylindrical magnets, which are constructed in a Halbach design and provide a constant 0.6~T magnetic dipole field in the hollow interior. The inside of the first $1.5~\m$-long magnet acts as a decay volume, and it is followed by an additional scintillator layer for timing and triggering. The remaining two $1.0~\m$-long magnets and three additional tracking stations form FASER's spectrometer. Located at the downstream end is the pre-shower station, consisting of two additional scintillator layers interleaved with tungsten plates, and, finally, the electromagnetic calorimeter with a depth of 25 radiation lengths.

Although the FASER main detector, composed entirely of electronic components, was optimized for long-lived particle searches, it is also able to detect neutrinos, as we will see. The signature of interest arises when a neutrino passes through the front veto scintillators and then interacts in the massive components at the front part of the detector, either the tungsten in FASER$\nu$ or the lead shield, producing an energetic hadronic shower. When this hadronic shower is produced near the back of the tungsten or in the lead shield, it is not contained, producing a distinctive signature of neutrinos in which no charged particles enter the detector and significant activity is recorded in the downstream electronic components of the detector. 

Although such a signal is indeed quite distinctive, there are nevertheless significant backgrounds that arise from the large number of energetic muons coming from the ATLAS IP. However, the different electronic detector components may also be used to separate the signal from these backgrounds: 
\begin{description}
\item [Scintillators] Neutrino interactions produce a large number of charged particles that activate the downstream scintillator layers but not the upstream front veto. In contrast, the vast majority of muons passing through the detector can be rejected using the front veto, leaving only a small number of events in which the muons pass through sides of the detector and barely miss the veto.  
\item [Tracker] In addition, the large number of charged particles produced in neutrino events can be seen in the tracking stations. Especially promising for this task is the IFT, which is located right behind the tungsten target. In contrast, muons typically deposit only a small amount of energy in tungsten, with $\langle dE/dx \rangle \sim 40~\mev/\cm$, and so no or only a small number of additional charged particles are expected to be present. 
\item [Calorimeter] Finally, the LHC neutrinos carry between several hundreds of GeV up to a few TeV of energy, with typically half of it being transferred to the hadronic shower. This can lead to a sizable energy deposit in the electromagnetic calorimeter, which is typically absent in the muon-induced background.
\end{description}

To illustrate these features, we show in \cref{fig:example_events} six example events. These events were obtained using the dedicated \texttt{FLUKA} simulation that we describe in \cref{sec:sim}.  They have passed the \textit{stringent} scintillator cuts that require no hits in the front veto scintillators and hits in all of the other scintillators. Each panel shows the distribution of charged particle hits in the IFT, as well as the energy deposit in the calorimeter. The top three panels show neutrino interactions occurring in the tungsten target. The event in the left panel contains a single track near the center of the tracker corresponding to a muon created during the neutrino charged current interaction. The absence of any further activity suggests that the scattering occurred in the upstream end of the tungsten target. The muon proceeds to generate a small $\sim200$ MeV deposit in the calorimeter. These sorts of neutrino events are the most difficult to distinguish from an incoming muon. In the middle and right panels we show two events in which the neutrino scattered closer to the end of the tungsten target. We observe a large number of hits that are strongly clustered and located in the center of the tracking station, accompanied by a sizable energy deposit in the calorimeter. 

The bottom three panels show muon events. In the left and middle panels, we show the most typical muon events that pass the stringent scintillator cuts; most muons either miss the first tracker completely or only pass near the edge of the tracker and deposit little to no energy in the calorimeter. In the right, we observe a rare muon event that generates a large number of hits at the edge of the tracker. Most of the shower seen in the tracker is stopped in the lead shield, and so there is very little energy deposited in the calorimeter.

In the rest of this study, we quantify these findings. In particular, we perform a dedicated \texttt{FLUKA} simulation of both neutrino and muon events in the FASER experiment. We use this to obtain the kinematic distributions, define observables, and develop an analysis strategy to distinguish the neutrino signal from background. 

\section{Simulation Setup}
\label{sec:muons}

\subsection{LHC Neutrinos and Muons}
\label{sec:flux}

\begin{figure*}[t]
  \centering
  \includegraphics[width=0.99\columnwidth]{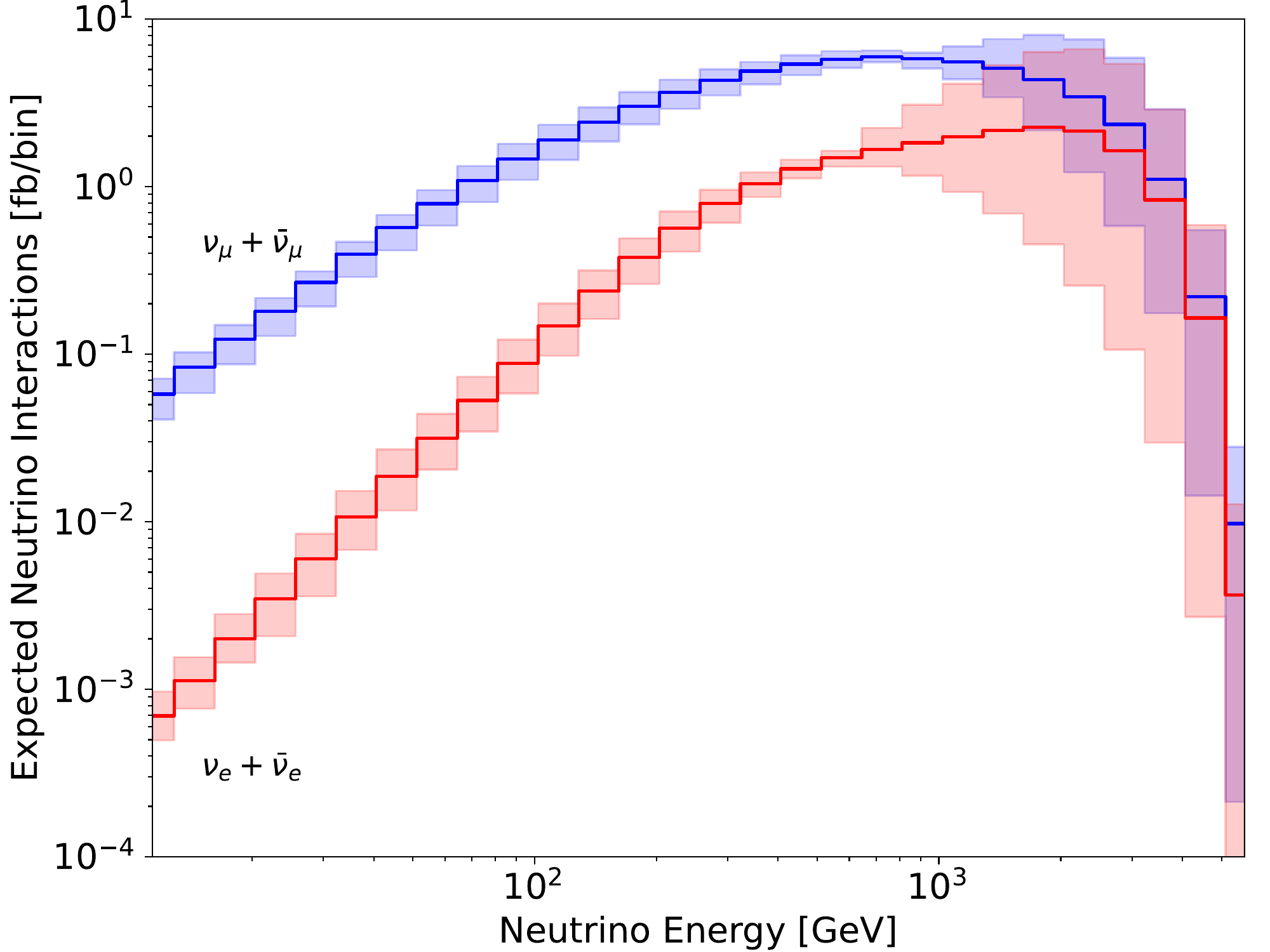}
  \includegraphics[width=0.99\columnwidth]{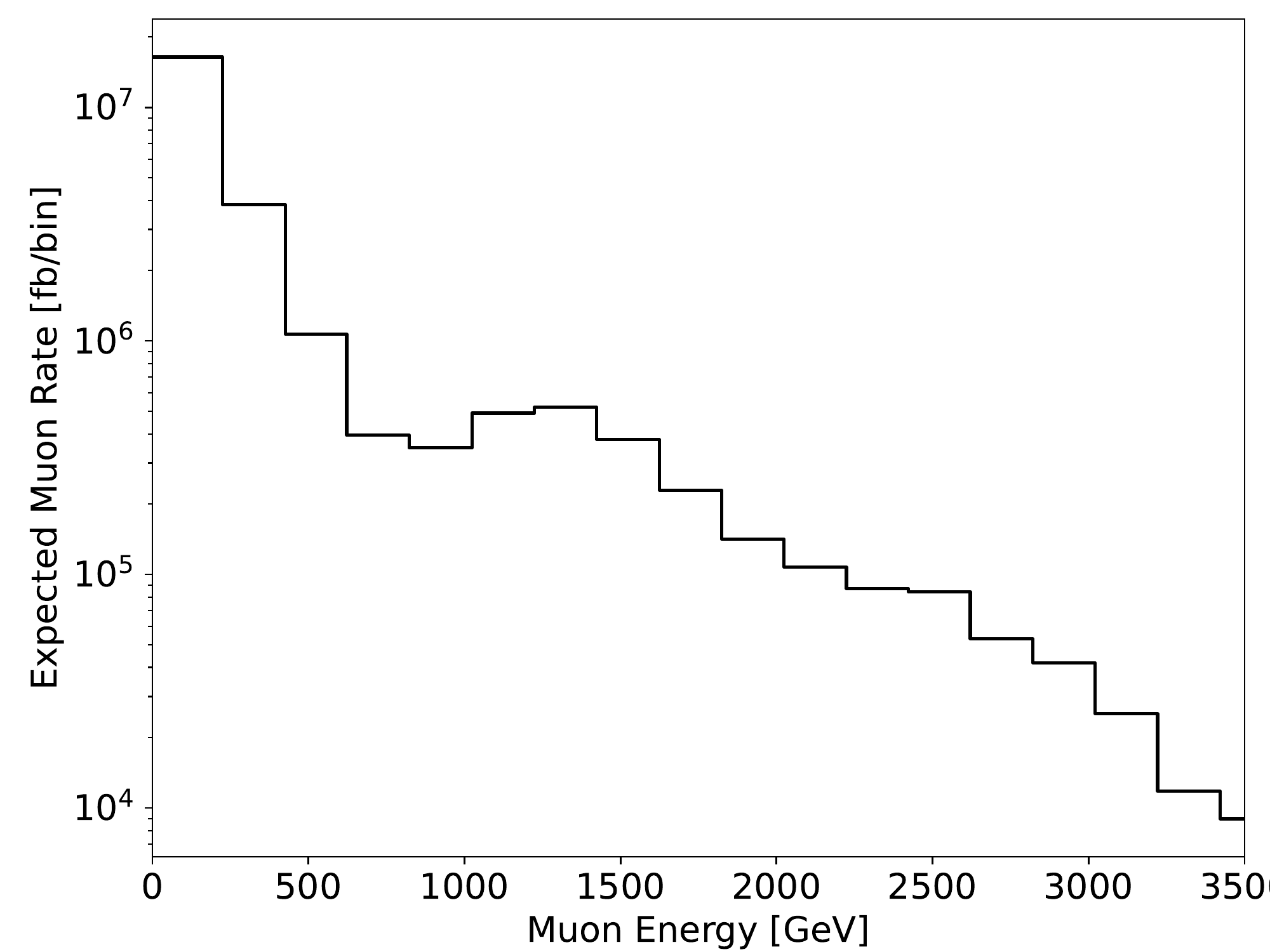}
\caption{{\bf Left:} The number of expected muon neutrinos (blue) and electron neutrinos (red) interacting with the target material as a function of their primary energy. The shaded region is a rough estimate of the flux uncertainty. {\bf Right:} The number of muons expected to evade the initial veto as a function of their primary energy.}
  \label{fig:counts_mu}
\end{figure*}

Before proceeding to the details of the simulation and analysis strategy, let us review the expected fluxes of the particles that pass through FASER.  For the signal, the relevant particles are the muon and electron neutrinos and anti-neutrinos.  For the background, the most relevant particles are muons produced near the ATLAS IP and muons and other particles produced in other ways by the collider, for example, through beam-gas interactions.

The neutrinos incident on FASER originate from forward hadrons produced at ATLAS, primarily pions, kaons and charmed hadrons. For this study, we use the neutrino fluxes presented in Ref.~\cite{Kling:2021gos}, which were obtained using a dedicated fast neutrino flux simulation to model the propagation and decay of long-lived hadrons in the forward LHC infrastructure. In particular, we use the central neutrino flux, which corresponds to an average of the predictions obtained using the event generators \texttt{Sibyll~2.3d}~\cite{Ahn:2009wx, Riehn:2015oba, Riehn:2017mfm, Fedynitch:2018cbl, Riehn:2019jet}, \texttt{EPOS-LHC}~\cite{Pierog:2013ria}, \texttt{QGSJET~II-04}~\cite{Ostapchenko:2010vb}, \texttt{DPMJET~III.2017.1}~\cite{Roesler:2000he, Fedynitch:2015kcn} and  \texttt{Pythia~8.2}~\cite{Sjostrand:2006za, Sjostrand:2014zea}.  To calculate the neutrino event rate, we use the neutrino scattering cross section on tungsten obtained using \texttt{Genie}~\cite{Andreopoulos:2009rq, Andreopoulos:2015wxa}. The resulting energy spectrum of interacting neutrinos, including both charged current and neutral current scattering in both the FASER$\nu$ tungsten target and the lead shield, is shown in the left panel of \cref{fig:counts_mu}. The average energy of the interacting neutrinos is $\mathcal{O}(\tev)$ for both $\nu_e$ and $\nu_{\mu}$. In addition to the central prediction, we also show a rough estimate of the neutrino flux uncertainty as a shaded band, which corresponds to the range of predictions obtained with the different generators. 

The dominant background to the considered neutrino signal is associated with LHC muons. These are produced at or near the ATLAS IP and pass through the roughly 100~m of rock and concrete before reaching FASER. The flux of LHC muons has been obtained by a dedicated \texttt{FLUKA} simulation performed by the EN-STI CERN group, which contains a realistic modelling of the LHC infrastructure and optics; it is presented in Ref.~\cite{FASER:2018bac}. The obtained muon energy spectrum is shown in the right panel of \cref{fig:counts_mu}, where we plot the expected rate for a $43 \times 43$ $\text{cm}^2$ muon beam evading the initial scintillator veto, which covers the central area of $30 \times 35~\cm^2$. In contrast to the spectrum of neutrinos that interact in FASER, which peaks near TeV energies, the muon flux peaks at low energies. The uncertainty associated with this flux estimate has been described as `a factor of a few' and predominantly originates from the limited simulation statistics~\cite{FASER:2018bac}.

In addition to the muons produced near the ATLAS IP, muons and other particles may be produced at other points along the LHC and arrive at FASER.  FASER is shielded from most of these particles by large amounts of rock and concrete.  An exception is particles produced in beam-gas collisions by Beam 1, which travels westward from LHCb past FASER on its way to ATLAS.  Particles produced by Beam 1-gas interactions can therefore travel up TI12 and pass through FASER without encountering any shielding.  This flux of particles has been observed, as discussed in Ref.~\cite{FASER:2018bac}.  However, in 2022, 6 $80\times 80 \times 80~\text{cm}^3$ concrete blocks were added at the base of TI12 to suppress this background, and this background can be further suppressed by the stringent scintillator cuts we discuss in \cref{sec:scintillators} and requiring that the scintillators be triggered with timing consistent with particles coming from the direction of ATLAS.  This background is therefore expected to be far below the dominant background of muons from ATLAS. 

\subsection{FLUKA Simulation}
\label{sec:sim}

We performed a Monte Carlo simulation of the signal and the main muon backgrounds using \texttt{FLUKA}~\cite{Battistoni:2015epi, FLUKA:new, FLUKA:web, Ferrari:2005zk, Bohlen:2014buj}. The \texttt{FLUKA} simulation is composed of the geometry of the apparatus, scoring (or recording) procedures, and the primary particles. 

The geometry specifies the details pertaining to the tungsten target and its aluminum cage, the lead shielding, the magnets and their fields, and the rock and concrete of the tunnel. A diagram of the setup generated by the \texttt{Flair}~\cite{Vlachoudis:2009qga, FLAIR:web} geometry viewer is shown in \cref{fig:layout}. The tungsten, aluminum, and lead are all assigned their corresponding default material in \texttt{FLUKA}, while the magnet, rock, and concrete are assigned custom materials that match their nuclear densities. The vacuum inside the magnets is filled with a 0.6 T magnetic field oriented toward the concrete floor, while the magnetic field inside the material is neglected.

For the scoring procedures, for each event, corresponding to the initialization of a single primary particle, we record a variety of data similar to the experimental observables with the \texttt{EVENTBIN} routine. 
\begin{description}
    \item [Scintillators] The scintillators are simulated as a volume of the \texttt{FLUKA}-defined \texttt{PLASCINT} material recording the energy deposited. The scintillators are $30\times30\times 2$ cm$^3$, except for the timing scintillator located between the magnets, which is $40\times 40 \times 1$ cm$^3$, and the front veto, which is $30\times 35\times 2$ cm$^3$. The scintillators located between the tungsten and lead shield are tilted 3.7$^\circ$ clock-wise and the scintillators located between the lead shield and the magnets are tilted 3.7$^\circ$ counter-clock-wise to match their orientation as installed in the FASER detector. We mark the scintillator as triggered during the event if the energy deposited exceeds 100 keV.
    \item [Tracker Stations] The trackers are simulated as $25 \times 25 \times 0.1$ cm$^3$ regions divided into 625 $1 \times 1 \times 0.1$ cm$^3$ bins with each bin scoring the number of charged hits in each bin.
    \item [Calorimeter] The calorimeter is simulated as a $24.3 \times 24.3 \times 13$ cm$^3$ (88 kg) lead target that has the same cross section and mass as the calorimeter in the FASER detector. The energy deposited in the calorimeter is recorded for each event.
\end{description}

To model the neutrino signal, we initialize electron and muon neutrino interactions evenly-distributed throughout the tungsten target and lead shielding with their momenta aligned with the long axis of FASER. Neutrinos in \texttt{FLUKA} interact immediately with the material they are initialized in, so the simulated neutrinos are weighted according to the expected number of neutrino interactions. The origin and spectra of the neutrino interactions are as discussed in \cref{sec:flux}. We simulate $3\times 10^5$ muon neutrino and $8\times 10^4$ electron neutrino interactions, while we expect only $10^4$ total neutrino interactions in all of Run 3. Due to oversampling of the neutrino interactions, the uncertainty in the signal rate predicted from the MC simulation is small, and our results are reliable. 

To model the background, we simulate the muon fluxes discussed in \cref{sec:flux}. We simulate muon samples in two regions in the transverse plane: a central region with area $30 \times 35$~cm$^2$, corresponding to the area covered by the front veto scintillators, and an outer region corresponding to a $43 \times 43$~cm$^2$ square centered on the beam collision axis, but omitting the $30 \times 35$~cm$^2$ region occupied by the front scintillators. In both cases, the primary muons start 16 cm in front of the initial scintillators with momenta aligned with the long axis of FASER, and we propagate them through 10 cm of rock.  The muon interactions in this rock can produce neutral hadrons before reaching the FASER detector. These neutral hadrons could be an important background, and they are included in the simulation. For the muon energy distribution, we divide the spectrum into the energy bins shown in \cref{fig:counts_mu}.  

For muons in the central region, we simulate approximately $10^4$ muons per energy bin. Given a conservative scintillator veto efficiency of 99.9\% for each scintillator (the expected efficiency is above 99.95\%), $\sim 10$ muons per $\ifb$ pass the central veto. These muons are, therefore, very well sampled in our simulation and are shown not to pose a problem.

For muons in the outer region, we simulate approximately $2 \times 10^6$ muons per energy bin. In all of Run 3, we expect $\sim 10^{7}$ muons in the outer region with energy above 1 TeV, which is computationally taxing to simulate. However, our analysis finds that only the $\sim 10^{6}$ high-energy muons with energy $E \gtrsim 1~\mathrm{TeV}$ are problematic, and these muons are sufficiently sampled in our simulation.  

\subsection{Cosmic Muons}
\label{sec:cosmic}

In addition to the muons and other particles produced by the collider, high-energy cosmic muons may also propagate to FASER. The flux and typical energies of cosmic muons are tiny compared to muons produced at the ATLAS IP.  On the other hand, cosmic muons may impinge on the FASER detector at significant angles relative to the beam collision axis.  One could therefore worry that they could more easily evade the front scintillator veto, but still deposit energy in the downstream components, thereby passing the stringent scintillator cuts and mimicking the signal with a greater efficiency than the LHC muons.

To investigate this, we have simulated the cosmic muons in \texttt{FLUKA}.  The flux of cosmic muons has been estimated by propagating the cosmic muon flux at the Earth's surface to the tunnel where FASER is located~\cite{DaveSavannah}.  It peaks for low energy muons coming from directly above, but muons coming at a large angle with respect to the beam axis will activate the scintillators in a way that is inconsistent with the timing signatures expected from neutrino events, and so can therefore be rejected. The most problematic muons are therefore those coming from near the direction of ATLAS, but the flux dramatically decreases for such angles and also for higher energies.   Over the 4 years 2022-2025 of Run~3, there are $\mathcal{O}(10^7)$ cosmic muons arriving at FASER from all angles and with energies above 10 GeV, but only  $\mathcal{O}(10^4)$ within an angle of $25^{\circ}$ of ATLAS and with energy above 500 GeV.

When these muons interact more than a few m before they reach the tunnel, the resulting showers are absorbed by the rock, but if they interact in a thin layer of rock that surrounds the tunnel, the resulting showers could propagate into the tunnel and trigger the scintillators.  We have modeled the tunnel as a cylinder with a radius of 2 m surrounded by rock, and with a concrete floor, as shown in Fig.~\ref{fig:layout}. The axis of the tunnel and FASER are offset by an angle of roughly \(17^{\circ}\).  We then simulate muons that start 2 m into the rock and consider primary muon trajectories that are pointed at all parts of the FASER/FASER$\nu$ detector.  We find that the expected rate of cosmic muon events that pass the stringent scintillator cut is $\mathcal{O}(10^{-3})$ events in the typical time it takes ATLAS to collect 1 $\ifb$.  This rate can be further suppressed by an order of magnitude by requiring that the muon arrive in coincidence with a bunch crossing.  

We conclude, then, that the cosmic muon rate is completely negligible and far below the LHC muon background rate.  Although the cosmic muon simulation could certainly be refined, this preliminary analysis indicates that it will be easily suppressed by the cuts that we impose to remove the LHC muons.  We note also that the cosmic muon background will be well-measured in a data-driven way when FASER is on, but the LHC beam is off. In the following, we therefore focus out attention on the dominant background from LHC muons. 

\section {Observables and Analysis \label{sec:observables}}

\begin{figure*}[t]
  \centering
  \includegraphics[width=0.49\textwidth]{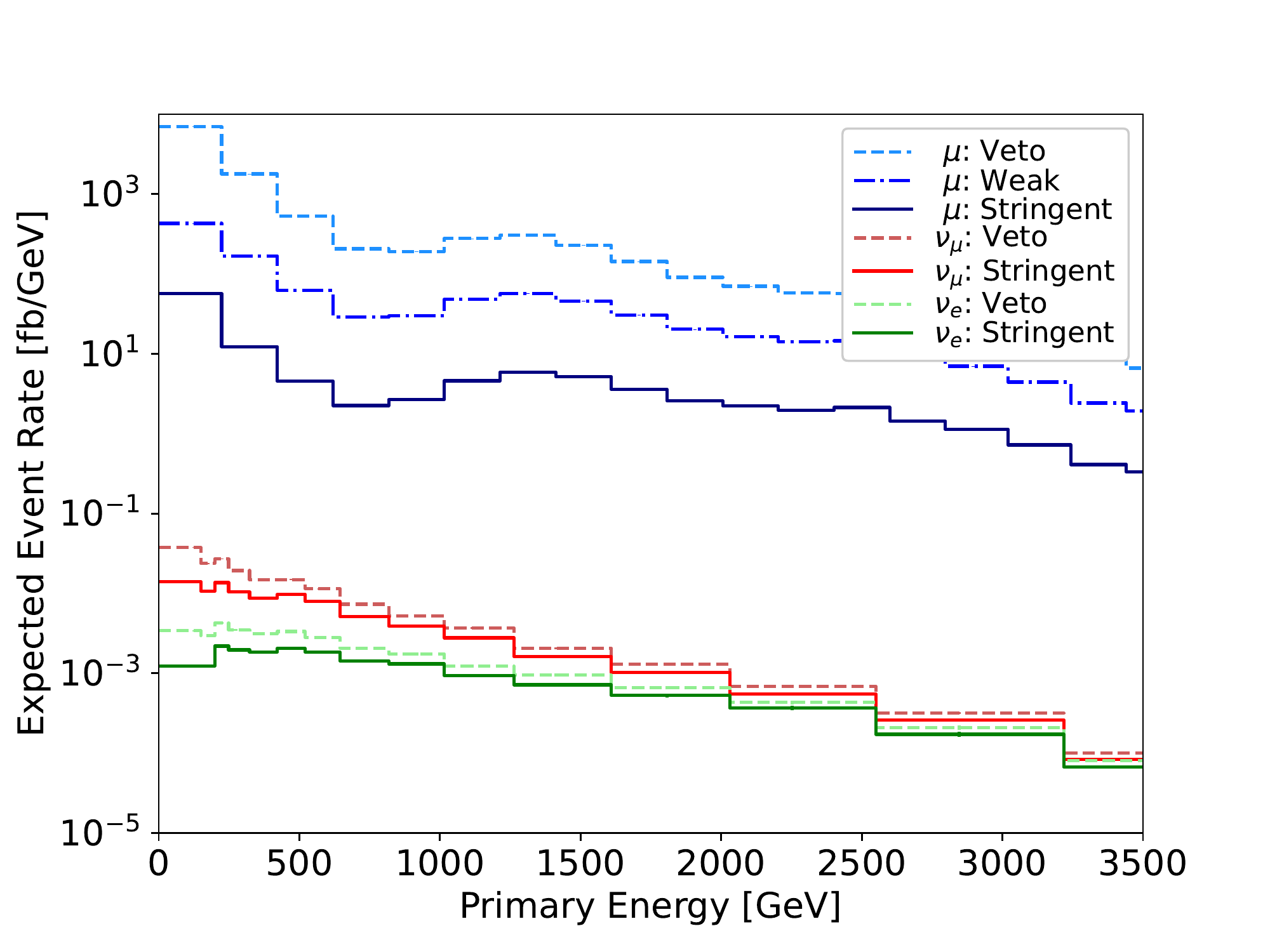}
  \includegraphics[width=0.49\textwidth]{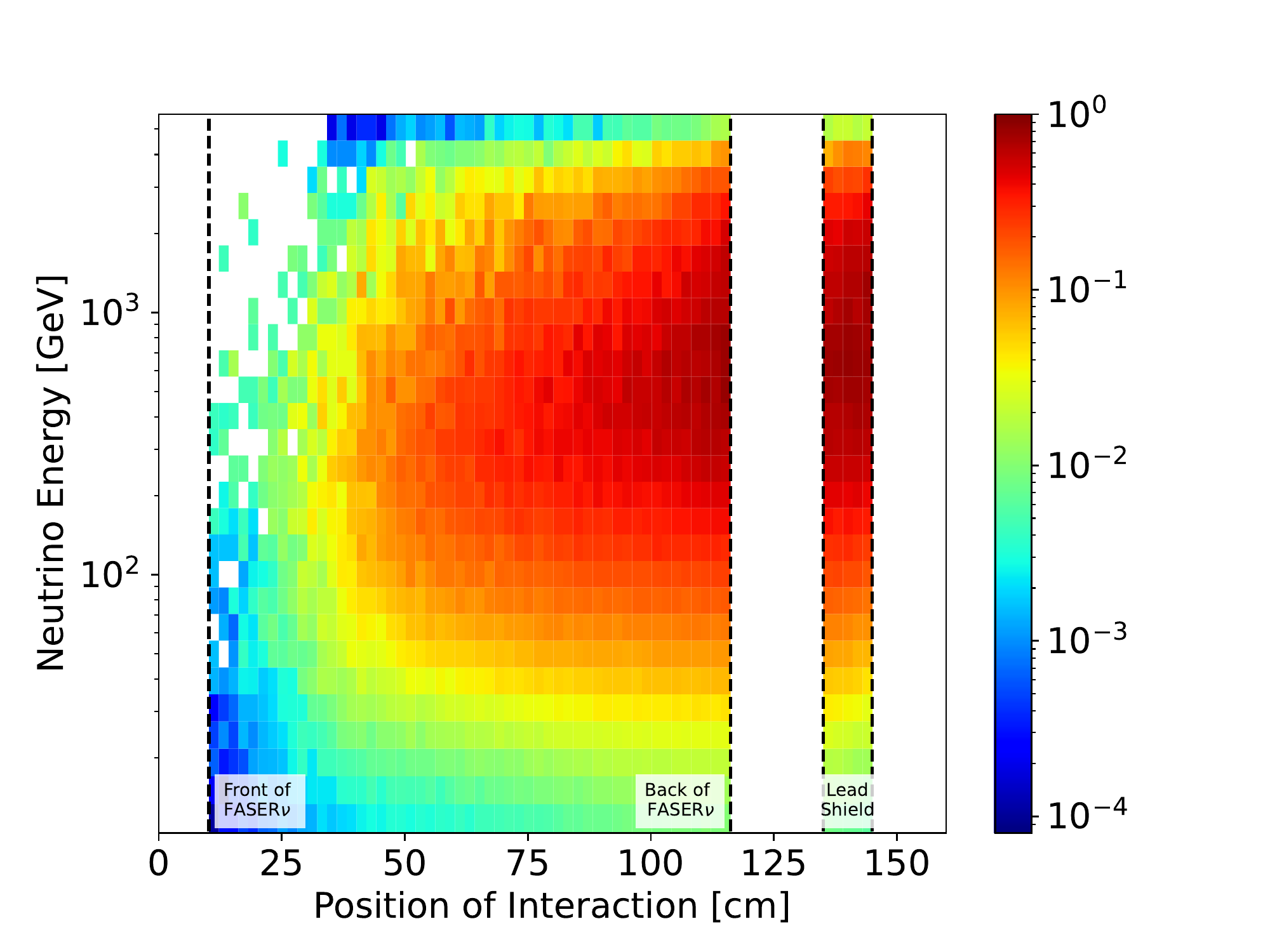}
  \caption{{\bf Left:} The expected rate of $\mu, \nu_\mu, \nu_e$ events passing various scintillator cuts as a function of the primary energy. The three cuts are described in \cref{eqn:scintcuts}. The \textit{stringent cut} reduces the background event rate by $\sim 10^2$ while keeping $\mathcal O(1)$ of the signal. {\bf Right:} The distribution of neutrino events that pass the stringent scintillator cut. The bins are colored according to the number of interactions expected to generate the signal for $10~\ifb$ integrated luminosity at ATLAS. The energy spectrum matches that of the neutrino interactions. The scintillator cut favors neutrino interactions at the back of the tungsten and in the lead shield.
  }
  \label{fig:scintcuts}
\end{figure*}

In the previous section, we described the experimental input for our analysis. In this section, we discuss how this may be used to separate the signal events from neutrino interactions from the background events arising from muons originating in the LHC.
In \cref{sec:scintillators} we describe how the majority of the muon background can be rejected using the scintillator activation pattern, in \cref{sec:calorimeter} we discuss the origin of large calorimeter energies and their rates for signal and background, and in \cref{sec:tracker_obs} we consider physically motivated tracker image observables. 

\subsection{Scintillators} \label{sec:scintillators}

Throughout FASER, there are nine scintillators that will trigger when a charged particle passes through with efficiencies that have been measured to above $99.95\%$. A striking feature of a muon neutrino interacting in FASER$\nu$ is the resulting muon which passes through the entirety of FASER. The muon from this interaction will proceed to trigger the scintillators following the interaction, but the background muons from cosmic rays and the LHC are {\em a priori} capable of producing the same signal. The two scintillators at the front of FASER provide an efficient veto for the majority of muons entering from the LHC. However, there remain $\sim 10^7$ muons that pass the edges of the initial veto for just $1~ \mathrm{fb}^{-1}$ at the LHC. Since there are expected to be on the order of $10-100$ neutrino interactions for the same integrated luminosity, this large flux of muons can easily generate backgrounds that eclipse the neutrino signal despite a low probability for an individual muon to generate a given signal. 

We considered many possible combinations of scintillator cuts and define three representative combinations:
\begin{equation} 
    \begin{array}{ c | c c | c c | c c | c | c c}
        \mathrm{Scintillators} & 1 & 2 & 3 & 4 & 5 & 6 & 7 & 8 & 9 \\ \hline
        \mathrm{only~veto} & \scintoff & \scintoff & \scintagn & \scintagn & \scintagn &  \scintagn & \scintagn & \scintagn & \scintagn \\
        \mathrm{weak~cut} & \scintoff & \scintoff & \scinton & \scinton & \scinton &  \scinton & \scintagn & \scintagn & \scintagn \\
        \mathrm{stringent~cut} & \scintoff & \scintoff & \scinton & \scinton & \scinton &  \scinton & \scinton & \scinton & \scinton 
    \end{array}
\label{eqn:scintcuts}
\end{equation}
The scintillators are numbered from the front of the detector (see \cref{fig:layout}): scintillators 1 and 2 are the veto scintillators at the front of the detector, 3 and 4 are just before the lead shield, 5 and 6 are just behind the lead shield, 7 is the timing scintillator just behind the first magnet, and 8 and 9 are part of the preshower.  In \cref{eqn:scintcuts} the $\scintoff$ indicate that the scintillator is off, the $\scinton$ indicate that the scintillator is on, and the $\scintagn$ indicate that the scintillator can be either on or off. 

\begin{figure*}[t]
  \centering
  \includegraphics[width=0.49\textwidth]{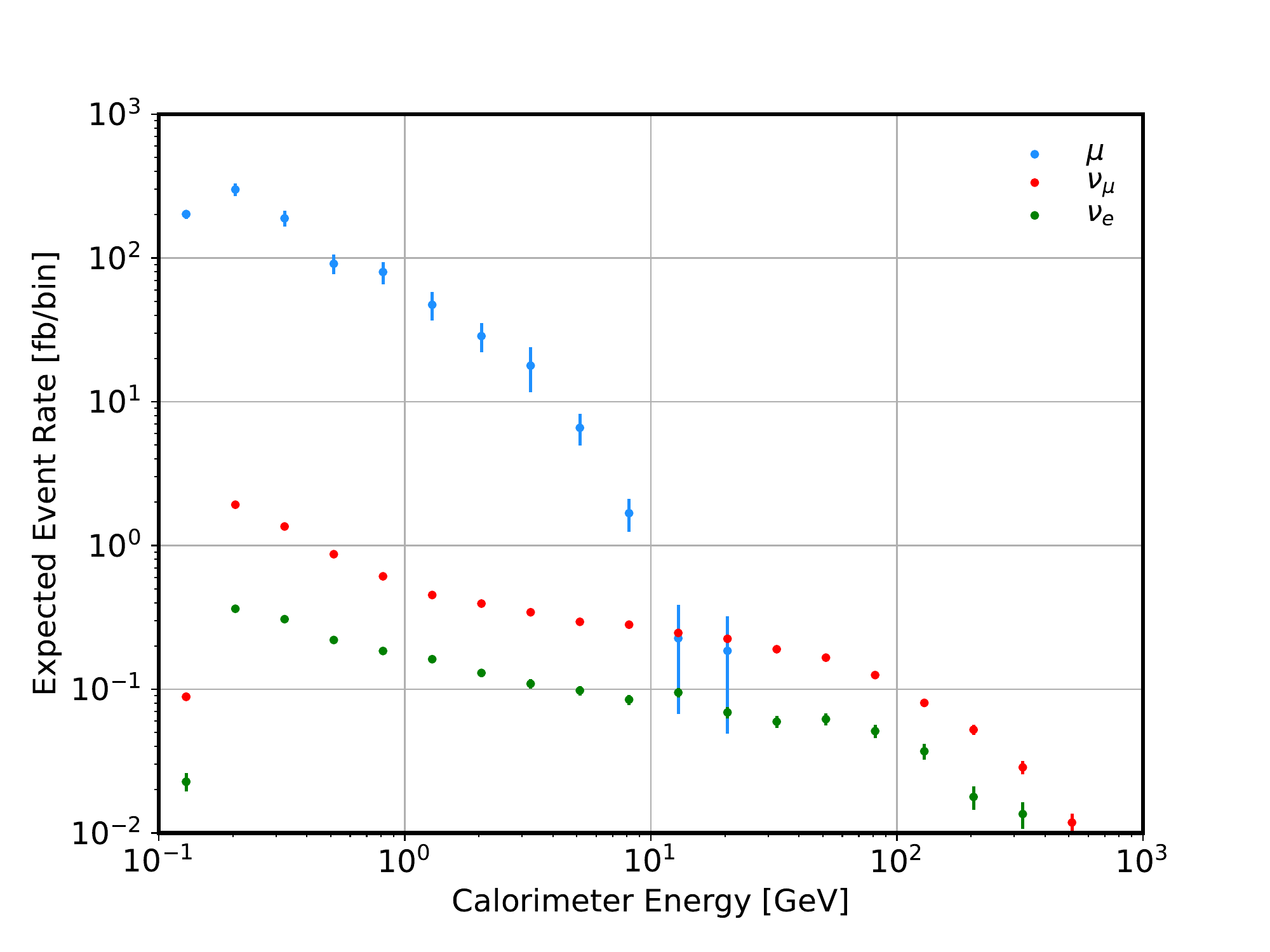}
  \includegraphics[width=0.49\textwidth]{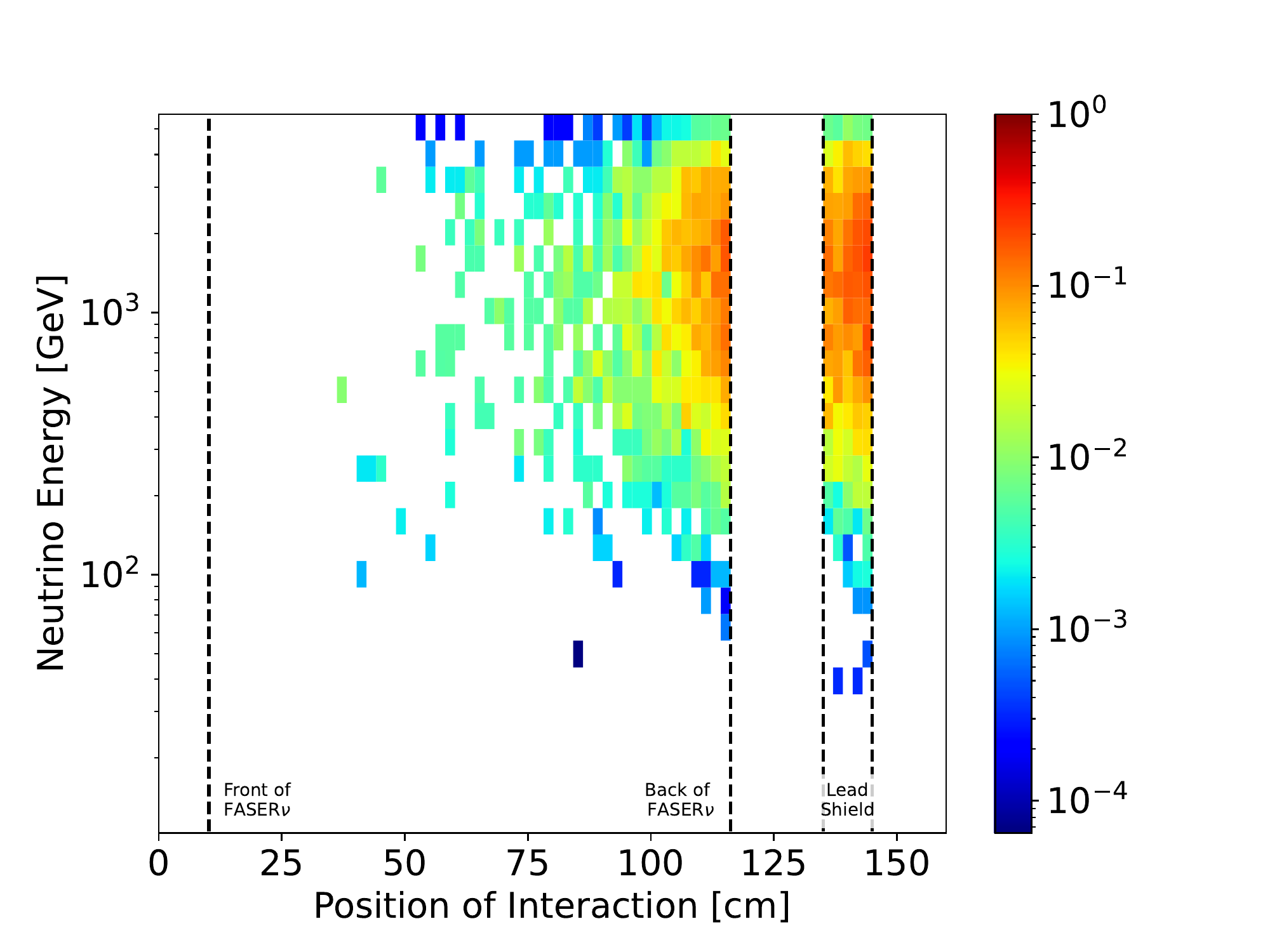}
  \caption{{\bf Left:} The expected event rate for $\mu, \nu_\mu, \nu_e$ as a function of the calorimeter energy assuming the stringent scintillator cut in \cref{eqn:scintcuts}. The error bars correspond to uncertainties from our MC statistics. The event rate for muons in the central region is separated from the muons in the outer annular region to distinguish between the origins of the muon background. {\bf Right:} The distribution of neutrino events that pass the stringent scintillator requirement and deposit at least 15 GeV in the calorimeter. The bins are colored according to the number of interactions expected to generate the signal for $10~\ifb$ integrated luminosity at ATLAS. The energy spectrum matches that of the neutrino interactions. The calorimeter energy cut favors high energy neutrinos interacting in the back side of the tungsten and in the lead shield.}
  \label{fig:ecalo}
\end{figure*}

In the left panel of \cref{fig:scintcuts}, we plot the expected event rate of muons, muon neutrinos, and electron neutrinos passing these three cuts as a function of their primary energy. As we can see, muons that barely miss the front veto in some cases still activate the downstream scintillators. The signal to background ratio improves from $\sim 10^{-7}$ when only applying a front veto cut to $\sim 10^{-4}$ with the stringent scintillator requirements. However, even with the drastic improvement in signal to background ratio provided by the most stringent cut, the scintillators are not sufficient to distinguish the background from signal alone. Combined measurements, from the calorimeter and interface trackers, are necessary to distinguish signal and background. Given the effectiveness of the stringent scintillator cut, it is assumed throughout the rest of our analysis.

In the right panel of \cref{fig:scintcuts}, we show the origins of the interacting neutrinos that pass the stringent scintillator requirement. The neutrino interactions are predominantly located at the back side of the tungsten or in the lead shield because the shower from the neutrino interaction near the front of the tungsten can ``backsplash'' and activate the initial veto. 

\subsection{Calorimeter Energy} \label{sec:calorimeter}

The FASER calorimeter is primarily designed to measure large energy deposits of the order of hundreds of GeV and above, which are the energies expected to be deposited by the decays of dark photons and other long-lived particles.  In addition, it can also be used to measure more moderate energy deposits resulting from neutrino interactions occurring in the front of the detector. Given the focus on high energies, however, the performance of the calorimeter may not be optimized for low energy deposits, however, and the calorimeter's ability to detect small energy deposits below $\mathcal{O}(10~\gev)$ may be limited. In the following, we will consider even low energy deposits in the calorimeter to be observable, but we note that it might not be possible to measure very small values below $\mathcal{O}(10~\gev)$.

Depending on the incident particles, the energy deposits can be very different:
\begin{description}
\item [Muons] As a minimum ionizing particle, the muon deposits on average $1.66~\mathrm{MeV/cm}$ in water~\cite{ParticleDataGroup:2020ssz}. A muon aligned with the beam axis travels through 13 cm of lead, depositing $\sim$147 MeV into the calorimeter. Of course, on rare occasions, a muon may also have a hard interaction and deposit more energy in the calorimeter; this is included in the simulation.  
\item [Hadrons] Charged pions, kaons, and other similar hadrons begin showering in the calorimeter, but most of the hadronic shower will escape the back of the calorimeter. 
\item [Electromagnetic Showers] Electrons and photons quickly shower and deposit most of their energy in the calorimeter. Therefore, they are the dominant source of large energy deposits. However, note that low-energy electrons entering the FASER decay volume are typically deflected by FASER magnets before reaching the calorimeter. 
\end{description}
Incident neutrinos are more likely to leave large energy deposits in the calorimeter than the incident muons. The high energy neutrino-nucleon interactions in the back of the tungsten target or lead shielding create showers of high energy hadrons and photons. In contrast, the incident muons will pass through FASER, leaving little trace apart from their ionizing track and emission of low energy photons through bremsstrahlung. 

This can be seen in \cref{fig:ecalo} where we plot the event rate for muons, muon neutrinos, and electron neutrinos as a function of the energy they deposit in the calorimeter. Low energy deposits are dominated by muon ionization as can be seen in the large jump in the event rate between 100 and 200 MeV. The muon event rate drops significantly at high energy deposits, while the neutrino rate remains largely intact. The neutrino event rate eventually surpasses the muon event rate around $\sim 10$ GeV, but, as we show in \cref{sec:results}, even calorimeter energy cuts as low as 1 GeV can significantly improve the discovery potential. 

Additionally, we plot the locations and primary energies of the neutrino events that deposit at least 15 GeV in the calorimeter. The events with high energy deposits are dramatically favored to result from neutrinos with energies $\sim$1 TeV which interact in the back 10 cm of the tungsten or in the lead shield.

The signal to background ratio can be quite high for cuts that require large energy deposit in the calorimeter, but the trade-off is a significant reduction in the signal event rate. As we show in the next section, tracker observables, either alone or in conjunction with the calorimeter energy, can be used to dramatically improve the discovery potential of our analysis while keeping a large fraction of the signal event rate. 

\subsection{Tracker Observables} \label{sec:tracker_obs}

\begin{figure*}[t]
  \centering
  \includegraphics[width=0.49\textwidth]{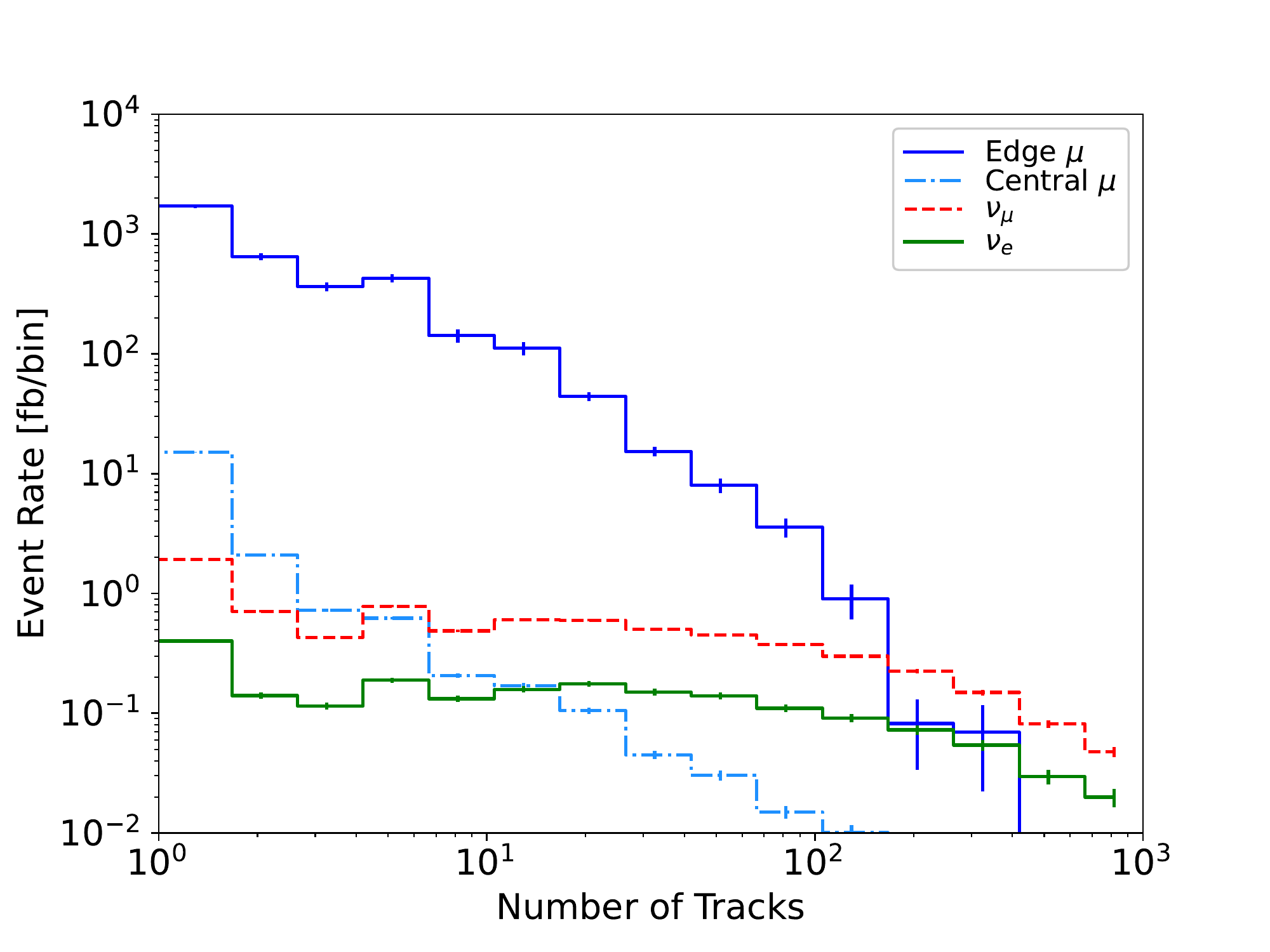}
  \includegraphics[width=0.49\textwidth]{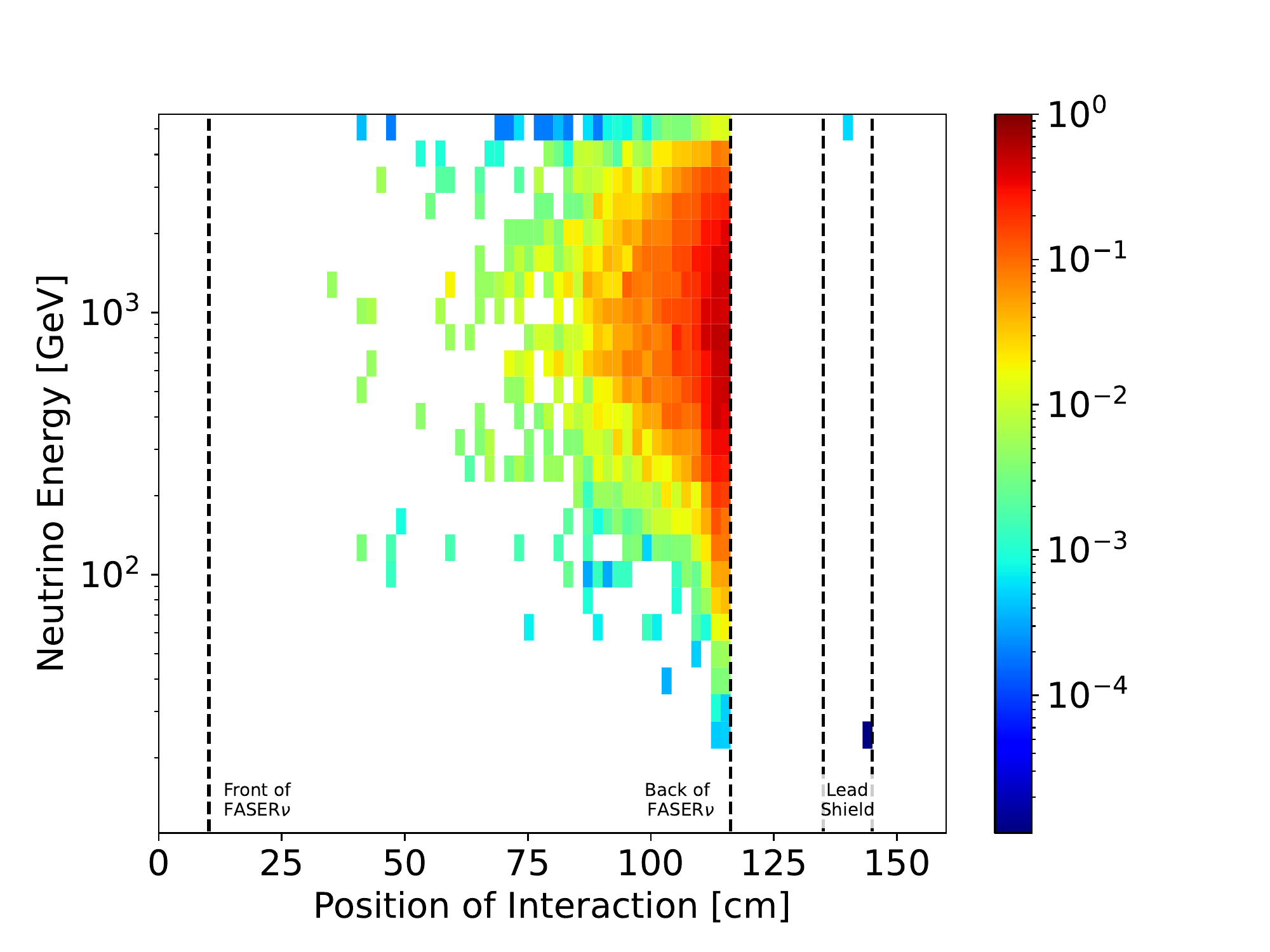}
  \caption{{\bf Left:} The expected event rate for $\mu, \nu_\mu, \nu_e$ as a function of the number of tracks in the first tracking station (IFT) assuming the stringent scintillator cut in \cref{eqn:scintcuts}. The error bars correspond to uncertainties from our MC statistics. The event rate for muons in the central region is separated from the muons in the outer annular region to distinguish between the origins of the muon background. {\bf Right:} The location of neutrino interactions that pass the stringent scintillator requirement with at least 50 charged tracks in the IFT. The bins are colored according to the number of interactions expected to generate the signal for $10~\ifb$ integrated luminosity at ATLAS. The energy spectrum matches that of the neutrino interactions. Increasing the number of tracks in the cut further favors interactions in the back of the tungsten.
  }
  \label{fig:ntracks}
\end{figure*}

There are four tracking stations that provide high resolution images of events as they progress through FASER. Each tracking station is equipped with SCT modules consisting of two pairs of silicon strip detectors with an $80~\mu m$ pitch size. They therefore allow one to identify the position of individual hits with excellent precision, and we therefore proceed in our analysis using the truth level information provided by \texttt{FLUKA}. We note, however, that their performance will be reduced when the number of hits becomes so large that almost all strips are activated. This effectively sets an upper limit on the number of observable hits of around $100/\cm^2$. A full simulation of the actual tracker would be needed to study the tracker performance at high track multiplicities and confirm the results derived here.

The most useful input is provided by the first tracking station, the IFT. Some example event displays have been shown in \cref{fig:example_events}. Conceptually, distinguishing between the neutrino signal and muon background can be seen as an image/pattern recognition problem, and there are a variety of modern techniques for this task. Instead, we take a different approach and define three physics-driven observables and focus on the first tracker located directly after the emulsion detector. While this approach helps to understand the physical differences between neutrino and muon interactions, an analysis using the full images of all trackers will undoubtedly perform better at distinguishing signal and background events.

\begin{figure*}[t]
  \centering
  \includegraphics[width=0.99\textwidth]{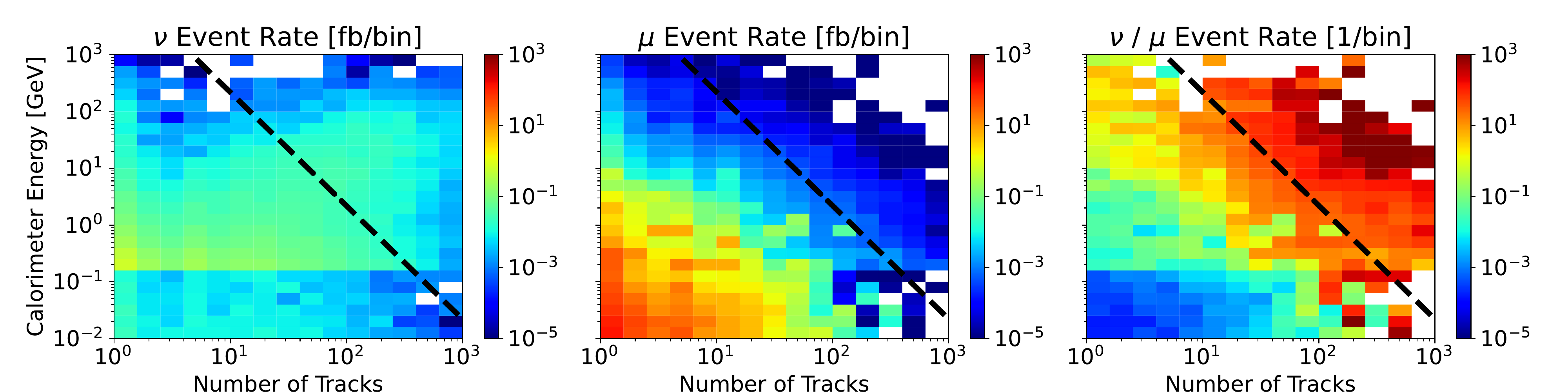}
  \includegraphics[width=0.99\textwidth]{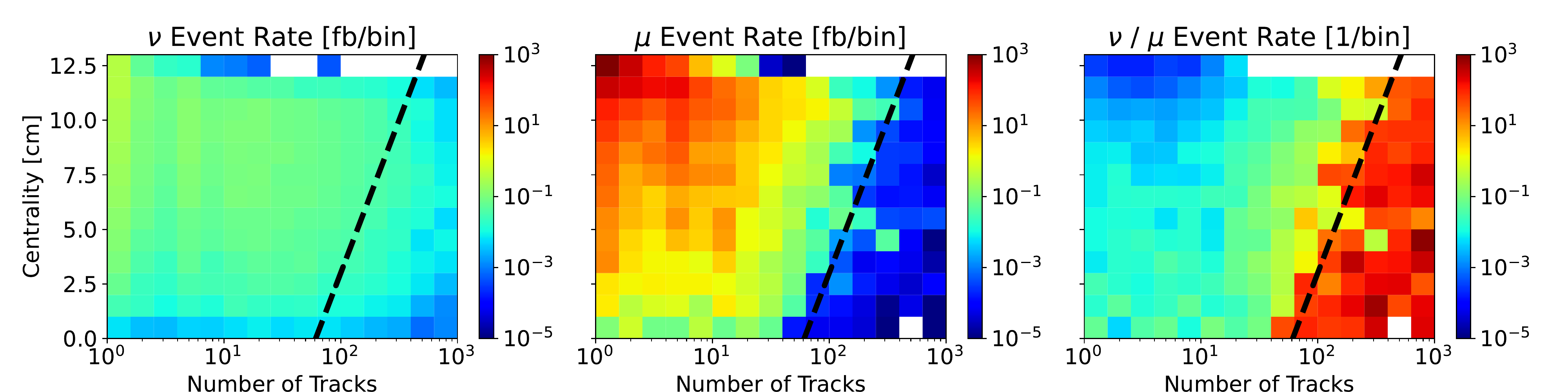}
  \includegraphics[width=0.99\textwidth]{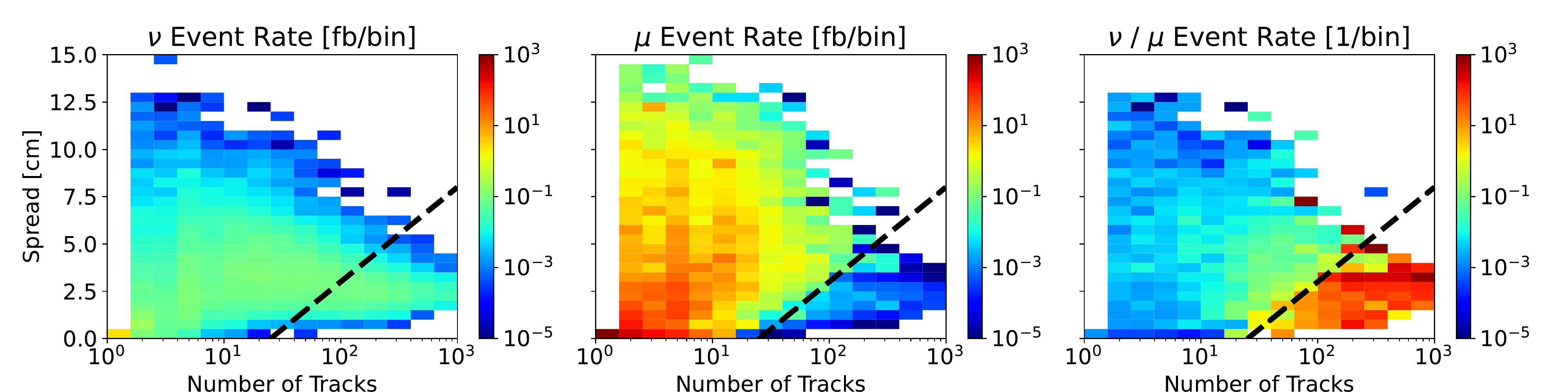}
  \caption{Distribution of events in calorimeter energy vs. track multiplicity (top row), centrality vs. track multiplicity (middle row) and spread vs. track multiplicity (bottom row). The rates for neutrinos and muons are shown in the left and center columns, respectively. The right columns shows the ratio of the neutrino and muon event rates, with red colors indicating promising regions for distinguishing signal from background. The dashed black lines correspond to the correlated cut discussed in the text.}
  \label{fig:2D_cut}
\end{figure*}

The track multiplicity $N$ is defined as the total number of tracks in each image. It is calculated as
\begin{equation}
    N = \sum_{i} n_i \ ,
\end{equation}
where $n_i$ is the number of tracks estimated in each pixel of the tracker image. The expected counts for muon and neutrino events as a function of the number of charged tracks in the first tracker are shown in the left panel of \cref{fig:ntracks}.  It is clear that a large number of tracks is a good indicator of a neutrino event due to the nature of the high energy neutrino-nucleus interaction. The neutrino interactions occurring in the back of the emulsion detector typically create an energetic hadronic shower containing large numbers of charged tracks. Meanwhile incident muons travel through the emulsion leaving a track possibly surrounded by a few ionized electrons. The primary way for muons to generate large numbers of charged tracks is via dramatic energy loss events, for example, via bremsstrahlung. The resulting high energy photon would then cause an electromagnetic shower containing a large number of electron tracks.

Since the primary mode for muons to generate large numbers of tracks is through an electromagnetic shower, most of the tracks going through the IFT will either be stopped by the lead shielding or diverted before the calorimeter by the magnets. In contrast, neutrino events often contain energetic hadrons that could be able to propagate until the calorimeter and deposit energy there. Thus large calorimeter deposits in events with large numbers of tracks should further distinguish neutrino signal events from muon background events. We present the neutrino and muon event rates as a function of the number of tracks and calorimeter energy in the left and central panels of the top row of \cref{fig:2D_cut}, respectively. The right panel shows the signal to background ratio. Indeed, the high signal to background region in the upper-right portion of the phase space, highlighted by a dashed line, indicates a potentially powerful search strategy. We will discuss the discovery potential of this region in \cref{sec:results}.

To further characterize the tracker images, we define two quantities, centrality and spread, which characterize the central position and width of the shower that the tracker images capture. We define the central position of the tracker image as
\begin{equation}
    (X,Y) = \frac{1}{N} \sum_i n_i \times (x_i, y_i)\,,
\end{equation}
where $(x_i, y_i)$ are the coordinates and $n_i$ are the number of tracks of each pixel.  The centrality $C$ is then the maximum of the two average coordinates
\begin{equation}
    C = \max(|X|,|Y|)\,.
\end{equation}
The intent is to quantify how close the event is to the center of FASER, which is aligned with the beam collision axis. Equivalently, this observable also quantifies the distance from the edge of the tracker plane. As the muon background predominantly originates from muons which pass the edges of the initial scintillator veto, centrality is physically motivated to distinguish between muons and neutrinos. 

We find that centrality alone is not sufficient to identify neutrino signal, but correlated cuts using the centrality can improve event selection. We show the two dimensional distribution of events in terms of centrality and track multiplicity for neutrinos and muons in the left and center panels of the middle row in \cref{fig:2D_cut} respectively. The right panel shows the signal over background ratio. We can see that the track multiplicities provide the most useful handle to isolate the signal. At small numbers of tracks, low centrality favors neutrino interactions, but the event rate is low. At high numbers of tracks, the centrality can help identify either neutrino or muon events as can be seen by the correlated cut illustrated with a dashed, black line. We present the discovery potential of such a cut in \cref{sec:results}.

\begin{figure*}[t]
\centering
\includegraphics[trim={0 0.5cm 0 0}, clip, width=0.99\textwidth]{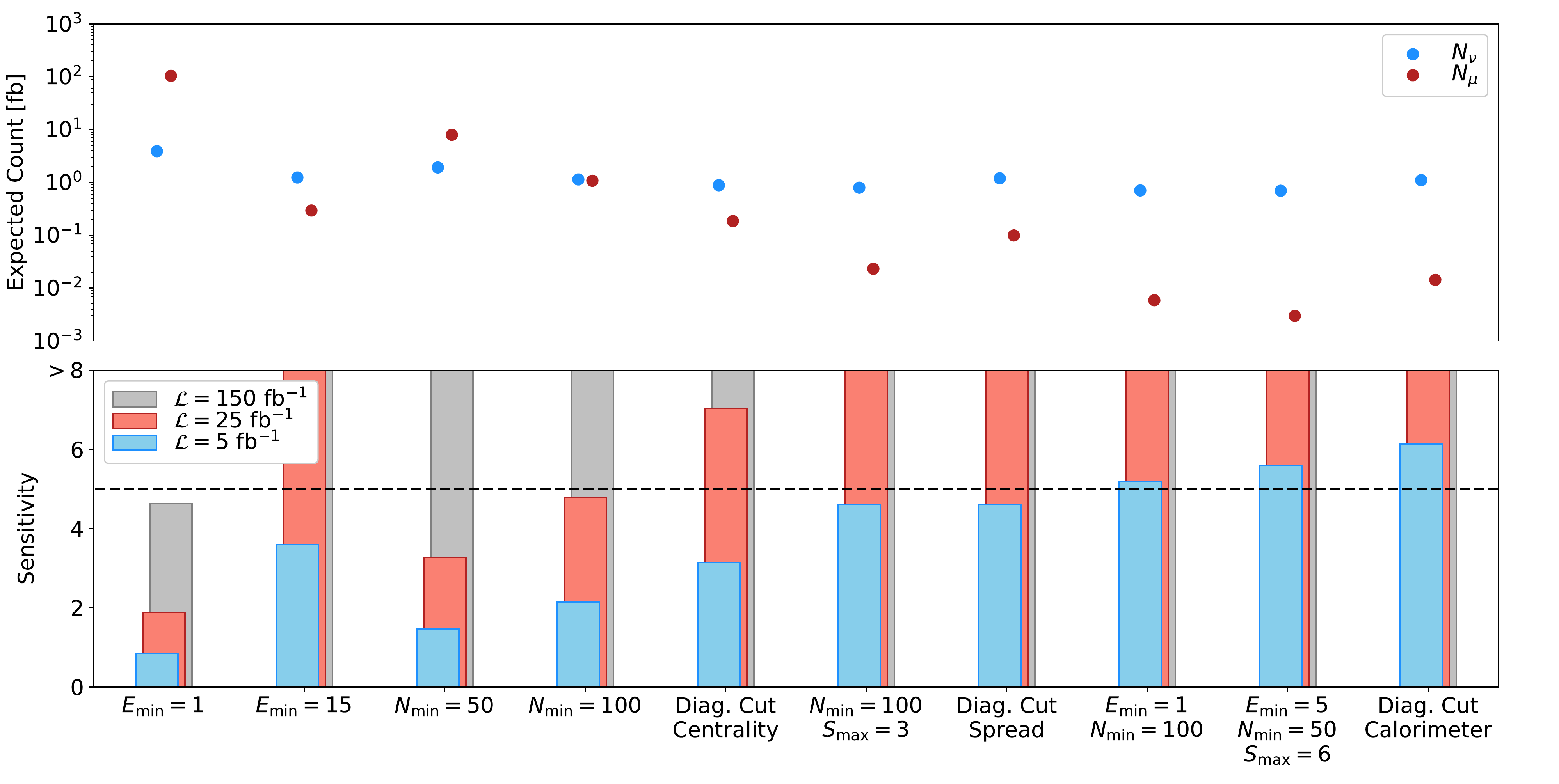}
\caption{In the upper plot, we show the expected number of $\mu$ and $\nu$ event rates for various cuts, and in the lower plot, we show the sensitivity for the same cuts for 5, 25, and $150~\mathrm{fb}^{-1}$. Here, $E_\mathrm{min}$ is the minimum required energy deposit in the calorimeter in GeV, $N_\mathrm{min}$ is the minimum required number of charged tracks in the IFT, and $S_\mathrm{max}$ is maximum allowed spread of the charged tracks in cm. The diagonal cuts in the 5th, 7th, and 10th columns corresponds to those indicated in the middle, bottom, and top rows of \cref{fig:2D_cut}, respectively. The stringent scintillator cut in \cref{eqn:scintcuts}, is required in all analyses. Note that in each cut, the number of neutrino events stays relatively constant, while the number of muon events changes drastically. We see that there is potential for a multivariate analysis to discover neutrino events at 5$\sigma$ with an integrated luminosity as low as $5~\mathrm{fb}^{-1}$.}
  \label{fig:cut_analysis}
\end{figure*}

The spread $S$ parameterizes the width of the shower seen in the IFT and is defined as 
\begin{equation}
    S = \left [ \frac{1}{N} \sum_i n_i \times \left ((x_i,y_i) - (X,Y) \right )^2 \right ]^{1/2}\,.
\end{equation}
As noted earlier, high-energy neutrino-nucleon interactions typically produce several showering particles that produce several tracks throughout the tracker whereas muons typically produce a few highly collimated tracks. While it is rare, muons can generate large numbers of tracks through an electromagnetic shower. The descendants of these shower will undergo several low energy interactions which cause the descendants to spread out around the muon track over short distances. In contrast, the large track numbers from neutrino events originate from a single high energy neutrino-nucleon interaction where the nuclear descendants travel longer distances between interactions, resulting in less spread out tracks than in the muon events.

Similar to centrality, spread alone is not capable of differentiating neutrino signal from muon background. However, there are easily identifiable regions of spread and number of tracks which can significantly improve the signal to background ratio. In the bottom row of \cref{fig:2D_cut}, we compare the neutrino and muon event rates as a function of spread and number of tracks to illustrate this point. The left and center panels show the neutrino and muon event rates, while the right panel shows the signal to background ratio. Just as with centrality, the signal to background ratio is largest at large numbers of tracks. The main difference with centrality is that the neutrino event rate remains large at high numbers of tracks with small spread. This can be seen by comparing the event rate of neutrinos under the dashed in lines in the bottom two rows of the left column of \cref{fig:2D_cut}. The additional handle of spread allows further rejection of muon background without removing the neutrino signal. In particular, the neutrino events favor small spread $\sim2-4$ cm for all numbers of tracks, while the muon event rate falls drastically at low spread and a high number of tracks, opening a promising search strategy. We present possible cuts and their discovery potential for various integrated luminosities in \cref{sec:results}.

\section{Results \label{sec:results}}

\begin{figure*}[t]
  \centering
  \includegraphics[width=0.99\textwidth]{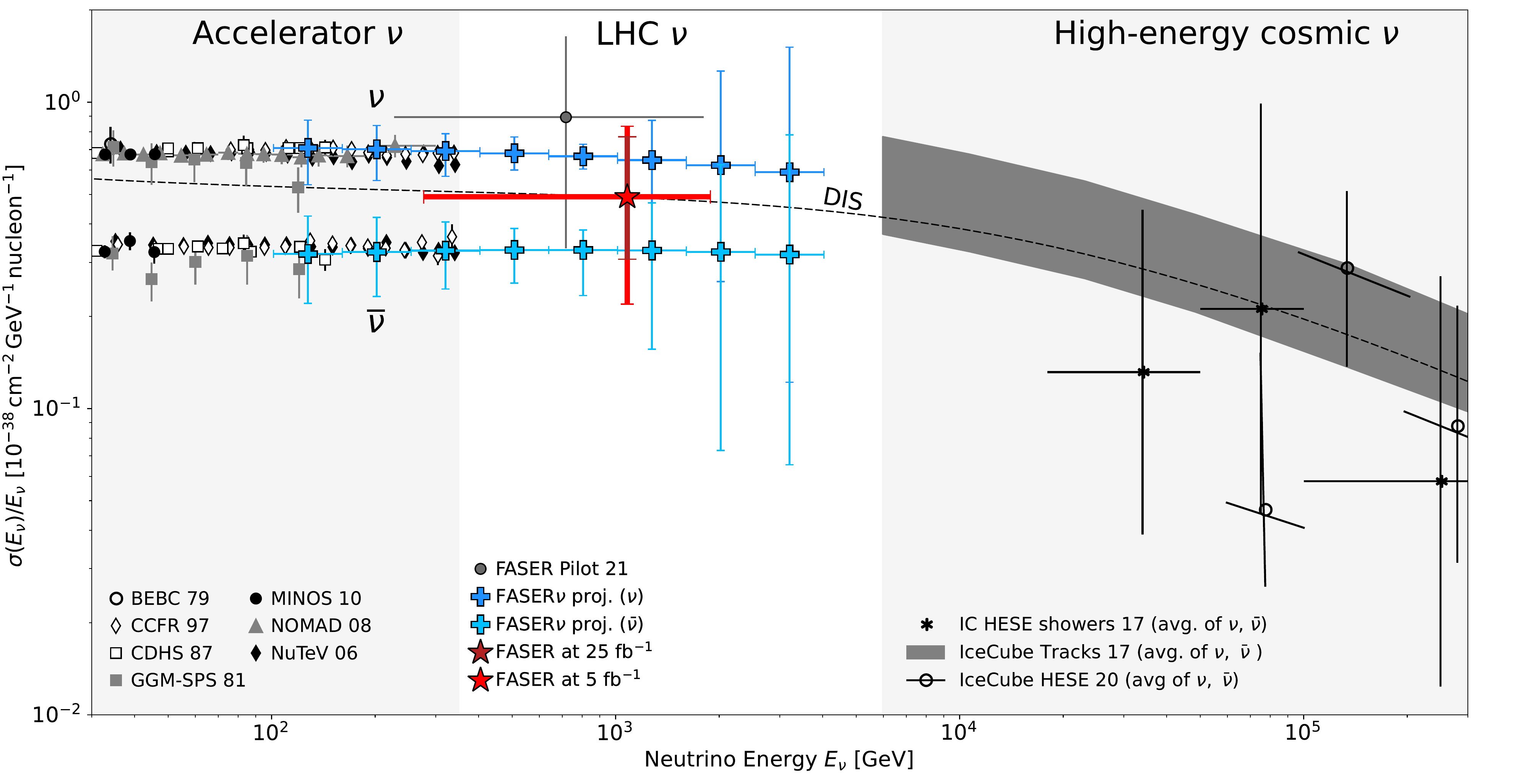}
  \caption{Measurements of the neutrino-nucleon charged current cross-section as a function of energy alongside the projected charged current cross-section measurements from FASER$\nu$ emulsion \cite{FASER:2019dxq} and the projected combined charged current and neutral current cross-section measurement of the analysis presented in this work.}
  \label{fig:cross_section}
\end{figure*}

In the previous section, we discussed the physics of several observables and their ability to distinguish neutrino events from muon events. Using these results, we design several possible analysis strategies to extract the neutrino signal. The corresponding results are shown in \cref{fig:cut_analysis}. Here the upper panel shows the number of neutrino signal and LHC muon background events after the selection cuts. Note that all of these cuts reduce the muon background by orders of magnitude while having little effect on the neutrino signal. The lower panel shows the corresponding expected statistical sensitivity, obtained from a likelihood ratio test, for several luminosities: $5~\ifb$, which roughly corresponds to the first month of data taking; $25~\ifb$, which roughly corresponds to luminosity expected in 2022; and $150~\ifb$, which is the nominal luminosity for LHC Run~3.

Starting on the left, we present four different single variable analyses requiring either a minimum energy deposit in the calorimeter, $E_\text{min}$, or minimum number of charged tracks in the IFT, $N_\text{min}$. We can see that stringent cuts $E_\text{min} = 2~\gev$ and $N_\text{min} = 100$ lead to a significance of about $3.5\sigma$ and $2\sigma$, respectively, at a luminosity of $5~\ifb$. Further improvements can be obtained using multivariate analyses. In particular, we present five different analyses, shown on the right, which are able to provide $\sim 5\sigma$ evidence for neutrinos already at a luminosity of $5~\ifb$. The cuts used in these approaches vary widely and rely on various subsets of experimental assumptions, implying a robustness to the claim that very little luminosity may be required for a $5\sigma$ discovery. \medskip

The sensitivity estimates discussed above only consider statistical uncertainties. However, in reality, several sources of systematic uncertainties will have to be taken into account for this analysis. The primary sources of uncertainty are expected to be related to particle fluxes.

Uncertainties associated with the muon flux will mainly effect analyses in which the number of muon-induced background events is comparable to or larger than the neutrino event rate. However, if the background rate can be reduced well below the signal, as in the last five examples presented in \cref{fig:cut_analysis}, flux uncertainties will have a minor impact. As described in \cref{sec:flux}, the muon flux and energy spectrum used in this study have been obtained using a dedicated \texttt{FLUKA} simulation, and the associated uncertainties are at a $\mathcal{O}(1)$ level. A first \textit{in-situ} measurement of the overall flux was performed in 2018 using an emulsion detector and a good agreement with the simulation was found. The situation will significantly improve in the near future once FASER starts to collect data. The muon flux and muon energy spectrum will therefore be constrained using the magnetized spectrometer in a data-driven way and significantly reduce the associated uncertainties. 

An additional uncertainty is associated with the neutrino flux, which will have a more direct impact on the expected sensitivity. A first quantitative estimate of this uncertainty was obtained in Ref.~\cite{Kling:2021gos} by comparing the predictions of different Monte Carlo event generators and shown as shaded band in \cref{fig:counts_mu}. The uncertainties are around tens of percent at lower neutrino energies but increase significantly at higher energies above $1~\tev$. This is due to an increasing contribution of neutrinos from charmed hadron decay to the overall flux for which the considered generator predictions differ by up to an order of magnitude. Dedicated efforts are needed, and have indeed already begun~\cite{Feng:2022inv, Bai:2020ukz, Jeong:2020idg, Bai:2021ira, Bai:2022jcs, Maciula:2020dxv}, to provide more reliable predictions for this forward charm production. \medskip

Although the observation of neutrinos at the LHC constitute an important milestone on its own, high energy neutrinos at the LHC also provide a array of opportunities for physics measurements. As a specific example, we interpret our proposed analysis as a measurement of the neutrino cross-section. This is illustrated in \cref{fig:cross_section} where we show the neutrino-nucleon interaction cross section as a function of the neutrino energy. Up to an energy of about 350~GeV, there are a variety of measurements of both the $\nu_\mu$ and $\bar \nu_\mu$ cross sections from accelerator neutrino experiments~\cite{Colley:1979rt, GargamelleSPS:1981hpd, Berge:1987zw, Seligman:1997fe, NuTeV:2005wsg, NOMAD:2007krq, MINOS:2009ugl}. In addition, there are several measurements using high energy cosmic neutrinos observed at IceCube which constrain the average $\nu_\mu + \bar \nu_\mu$ cross sections at energies between 10~TeV and 10~PeV~\cite{IceCube:2017roe, Bustamante:2017xuy, IceCube:2020rnc}. In the future, measurements with ultra-high energy cosmic neutrinos will allow to extend these measurements to even higher energies~\cite{Valera:2022ylt}.

At TeV energies, in between the accelerator neutrino and high-energy cosmic neutrino experiments, there remains a gap that has evaded cross section measurements. This gap can be accessed by LHC neutrinos. Indeed, the FASER$\nu$ pilot detector results can already be used to constrain this region, although with large uncertainties. We have reinterpreted the results presented in Ref.~\cite{FASER:2021mtu} and obtained the first measurement on the neutrino cross section at TeV energies. This is shown as a gray error bar in \cref{fig:cross_section} and takes into account both the uncertainty on the measured number of neutrinos as well as the flux uncertainty. Since the detector lacked the ability to identify leptons, this result should be understood as a constraint on the overall interaction strength for both charged and neutral current interactions of neutrinos and anti-neutrinos of all flavors. 

Shown in red we include the projected sensitivity that could be obtained with the analysis strategy presented in this work, assuming a luminosity of both $5~\ifb$ and $25~\ifb$. Similar to the FASER pilot run, this constrains the average $\nu_\mu$ and $\bar \nu_\mu$ cross section due to a lack of energy resolution and flavor identification abilities. The energy error bars are the 68\% CL in the energy of a neutrino interaction, and the cross-section error bars are the combined uncertainties from the statistics of the limited number of events and the neutrino flux uncertainties. We emphasize that, while the FASER pilot detector recorded the first neutrino interaction candidates at the LHC, the analysis we are proposing would be the first 5$\sigma$ signal of TeV neutrinos constraining the neutrino-nucleon cross-section in this novel region.

Finally, we also present the projected cross section sensitivity for FASER$\nu$ assuming a luminosity of $150~\ifb$. Unlike the analysis strategy proposed in this paper, the emulsion neutrino detector will provide additional information on the neutrino interaction which allows to both identify the leptons and estimate the neutrino energy~\cite{FASER:2019dxq}. In addition, the interface with the FASER spectrometer will measure the final state muon charge and therefore distinguish neutrinos and anti-neutrinos. Here we take into account a geometrical efficiency of 42\% for a muon produced in the emulsion detector to enter the smaller FASER spectrometer. The corresponding results are shown in blue and include both statistical and flux uncertainties. 

\section{Conclusions \label{sec:conclusions}}

The 2021 detection of far-forward neutrino candidates by an 11 kg FASER pilot detector~\cite{FASER:2021mtu} has signaled the opening of the new field of LHC neutrino physics. With the successful installation of the ton-scale detectors FASER$\nu$ and SND@LHC in the far-forward regions 480 m from the ATLAS IP, it is expected that $\sim 10,000$ TeV-scale neutrinos will be detected with the $150~\ifb$ of integrated luminosity at LHC Run 3 from 2022-25.

The full analysis of FASER$\nu$ and SND@LHC data, especially the emulsion data, will take time.  In this work, we have shown that, even without an analysis of the emulsion data, a $5\sigma$ discovery of collider neutrinos is possible with as little as $5~\ifb$ of integrated luminosity.  In addition, this electronic-detector-only analysis provides an alternative way of studying LHC neutrinos with experimental systematics that are very different from emulsion detectors. It therefore provides an independent cross check and an alternative view that may be sensitive to different new physics effects. Of course, as noted in \cref{sec:intro}, a detailed study with full simulation by the FASER Collaboration is needed to confirm the proposed analysis.  

The analysis is designed to isolate neutrinos that pass through the front veto scintillators and interact in the back of the tungsten target of the FASER$\nu$ detector or the lead shielding.  The resulting shower of particles may then be seen as charged tracks in the IFT and downstream trackers, in the downstream scintillators and through the deposit of significant energy in the calorimeter. The leading background is from muons produced at the LHC. Very rarely, these may pass through the front veto scintillators undetected, or they may just miss these scintillators, interact in the material on the sides of the detector, and produce particles that are detected in the downstream components. 

We have simulated the neutrino signal and muon-induced background in {\tt FLUKA}. We have found that the signal sensitivity is maximized by requiring a set of stringent scintillator cuts, in which there are no hits in the front veto scintillators, but hits in all of the other scintillators.  In addition to this requirement, we have examined the effect of requiring, in various combinations, a minimal number of charged tracks in the IFT, a maximal spatial spread of these tracks in the transverse plane, and a minimal energy deposit of 1, 5, or 15 GeV in the calorimeter.  The results are given in \cref{fig:cut_analysis}.  We see that the most effective set of cuts retain roughly 1-10\% of the neutrino signal rate, while simultaneously suppressing the background by many orders of magnitude.  A $5\sigma$ discovery is possible with the data collected in the early running of LHC Run 3 in 2022. The study of LHC neutrinos will therefore quickly pass through the discovery stage into the stage of studying TeV neutrinos.  As an example, in \cref{fig:cross_section} we show that neutrino detection at FASER with just $5~\ifb$ will provide an interesting constraint on the neutrino-nucleon cross section in the currently open window from $E_{\nu} \sim 350~\gev - 10~\tev$. 

Although our results are promising, we emphasize that our analysis is limited to using the number of tracks and spread of the first tracker image and simple cuts. A thorough analysis of the full tracker data of all four tracker stations could improve this analysis, allowing a discovery with less integrated luminosity at the LHC and a better measurement of the neutrino-nucleon cross section. Additionally, the event rate could deviate from SM predictions. Such an anomaly could be the sign of new muon, neutrino, or dark physics. An analysis of these scenarios is outside the scope of this study, but it would be interesting to consider these possibilities.

\section*{Acknowledgements}

We thank Mauricio Bustamante for providing neutrino cross section data displayed in \cref{fig:cross_section}. We also thank Aki Ariga, Tomoko Ariga, Jamie Boyd, Dave Casper, Hide Otono, Brian Petersen, and Savannah Shively for useful discussions. We are grateful to the authors and maintainers of many open-source software packages, including
\texttt{CRMC}~\cite{CRMC}, 
\texttt{Flair}~\cite{Vlachoudis:2009qga, FLAIR:web},
\texttt{FLUKA}~\cite{Battistoni:2015epi, FLUKA:new, FLUKA:web, Ferrari:2005zk, Bohlen:2014buj},
\texttt{Rivet}~\cite{Buckley:2010ar,Bierlich:2019rhm},
and \texttt{scikit-hep}~\cite{Rodrigues:2019nct, Rodrigues:2020syo}.
This work of J.A., J.L.F., and M.W.~is supported in part by U.S.~National Science Foundation (NSF) Grant PHY-1915005. The work of J.A. is supported in part by NSF Grant OMA-2016244. The work of J.L.F.~is supported in part by NSF Grant PHY-2111427, Simons Investigator Award \#376204, Simons Foundation Grant 623683, and Heising-Simons Foundation Grants 2019-1179 and 2020-1840.  The work of A.I.~is supported by the DOE under Grant No.~DE-SC0016013. The work of F.K.~is supported by Deutsche Forschungsgemeinschaft under Germany’s Excellence Strategy---EXC 2121 Quantum Universe---390833306.  The work of M.W.~is supported at the Technion by a Zuckerman Fellowship. 

\appendix


\bibliography{scintneut}

\end{document}